%% file: jpsito4l.tex
\documentclass[amsmath,amssymb,twocolumn,prl,floatfix,showpacs,nofootinbib]{revtex4-1}
\usepackage{amsmath}
\usepackage{color}
\usepackage{setspace}
\usepackage{overpic}
\usepackage{amssymb}
\usepackage{overpic,graphicx}
\usepackage{dcolumn}
\usepackage{epstopdf}
\usepackage{subfigure}
\usepackage{multirow}
\usepackage{ulem}
\usepackage[colorlinks,urlcolor=blue, citecolor=blue,linkcolor=blue] {hyperref}
\usepackage{lineno}
\usepackage{bm}
\usepackage{rotating}
\usepackage[utf8]{inputenc}
\newcommand{\psip}{\psi(3686)}

\newcommand{\Jll}{J/\psi \rightarrow \ell^{+}\ell^{-}}
\newcommand{\Jllll}{J/\psi \rightarrow \ell_1^{+}\ell_1^{-}\ell_2^{+}\ell_2^{-}}
\newcommand{\Jeeee}{J/\psi \rightarrow e^{+}e^{-}e^{+}e^{-}}
\newcommand{\Jeemm}{J/\psi \rightarrow e^{+}e^{-}\mu^{+}\mu^{-}}
\newcommand{\Jmmmm}{J/\psi \rightarrow \mu^{+}\mu^{-}\mu^{+}\mu^{-}}
\newcommand{\Pppj}{\psip \rightarrow \pi^{+}\pi^{-}J/\psi}
\newcommand{\Jgee}{J/\psi \rightarrow \gamma e^{+}e^{-}}
\newcommand{\Jgmm}{J/\psi \rightarrow \gamma \mu^{+}\mu^{-}}
\newcommand{\Jeepp}{J/\psi \rightarrow e^{+}e^{-}\pi^{+}\pi^{-}}
\newcommand{\Jpppp}{J/\psi \rightarrow \pi^{+}\pi^{-}\pi^{+}\pi^{-}}
\newcommand{\eegee}{e^{+}e^{-} \rightarrow \gamma e^{+}e^{-}}

\let\oldequation\equation
\let\oldendequation\endequation
\renewenvironment{equation}
 {\linenomathNonumbers\oldequation}
 {\oldendequation\endlinenomath}

\begin{document}
%\linenumbers

\title{\bf \boldmath
Observation of $J/\psi$ decays to $e^{+}e^{-}e^{+}e^{-}$ and $e^{+}e^{-}\mu^{+}\mu^{-}$
}

\input{BESIII_author}

\begin{abstract}
  Using a data sample of $4.481\times 10^8$ $\psip$ events collected with the BESIII detector, we report the first observation
  of the four-lepton-decays $J/\psi\to e^+e^-e^+e^-$ and $J/\psi\to e^+e^-\mu^+\mu^-$ utilizing the process $\psip\to \pi^+\pi^- J/\psi$.
  The branching fractions are determined to be $[5.48\pm0.31~(\rm stat)\pm0.45~(\rm syst)]\times 10^{-5}$ and $[3.53~
  \pm0.22~(\rm stat)\pm0.13~(\rm syst)]\times 10^{-5}$, respectively. The results are consistent with theoretical predictions. No significant signal is
  observed for $J/\psi\to \mu^+\mu^-\mu^+\mu^-$, and an upper limit on the branching fraction is set at $1.6\times 10^{-6}$ at
  the 90$\%$ confidence level. A CP asymmetry observable is constructed for the first two channels, which is measured to be $(-0.012\pm0.054\pm0.010)$ and $(0.062\pm0.059\pm0.006)$,
respectively. No evidence for CP violation is observed in this process.
\end{abstract}

%\pacs{13.20.Fc, 14.40.Lb}

\maketitle

\setlength{\oddsidemargin}{-0.6cm}
\setlength{\oddsidemargin}{-0.5cm} \addtolength{\topmargin}{10mm}
\hoffset -0.2 in

\oddsidemargin  -0.2cm
\evensidemargin -0.2cm

The standard model (SM)~\cite{Glashow:1959wxa,Salam:1959zz,Weinberg:1967tq} of particle physics has been remarkably successful
in explaining almost all experimental results in this field and has precisely predicted a wide variety of phenomena. However, it is clear that
the model is incomplete. Many questions still remain unanswered, such as the origin of neutrino masses, the mechanism behind
the observed matter-antimatter asymmetry, the nature of dark matter, \emph{etc}. These facts have prompted particle physicists
to put forward new theoretical frameworks, and to search experimentally for new physics beyond the SM. On the one
hand, lepton flavor universality (LFU) is expected to be obeyed in SM. In recent years, however, indications for violation of
LFU have been reported in semileptonic decays of the kind $b\to s~\ell^+\ell^-$. In 2014, LHCb measured the ratio of branching
fractions~(BFs) $R_K = {\cal B}(B^+\to K^+\mu^+\mu^-)/{\cal B}(B^+\to K^+ e^+e^-)$, and found a deviation from the SM prediction
by 2.6$\sigma$~\cite{Aaij:2014ora,Bordone:2016gaq}. The measurements have continuously been updated by LHCb~\cite{Aaij:2019wad}
and Belle~\cite{Abdesselam:2019lab}. Very recently, LHCb reported their latest result with full Run I and Run II data,
which deviates from SM prediction by more than $3\sigma$~\cite{Aaij:2021vac}. On the other hand, recently, the FNAL
Muon $g-2$ experiment~\cite{Abi:2021gix} released their first measurement of the anomalous magnetic moment of the muon.
The result confirmed the long-standing discrepancy with the SM prediction, which was previously observed by the E821 experiment at
Brookhaven National Laboratory (BNL) \cite{Bennett:2006fi}. After taking into account the contribution from BNL-E821, the deviation
between measurement and theoretical prediction amounts to around 4.2$\sigma$. These findings indicate there possibly exist
new particles or new couplings to leptons, especially for the muon~\cite{Ban:2021tos,Marzocca:2021azj,Borah:2021jzu,Lancierini:2021sdf, Du:2021zkq,Kawamura:2021ygg,Chen:2021vzk,Yin:2021yqy,Arcadi:2021cwg,Baum:2021qzx,Cadeddu:2021dqx,Ellis:2021ixr}. It is therefore urgent to investigate the validity of LFU in other experiments.

$\Jll$, where $\ell$ may be either $e$ or $\mu$, are two such precisely measured channels, and their measured BFs
are consistent with  Quantum Electrodynamics~(QED) calculations~\cite{bes3jpsi2ll}. Other purely leptonic decays, which have never been studied
experimentally, are $J/\psi \to \ell_1^+ \ell_1^- \ell_2^+ \ell_2^-$, where $\ell_1 = \ell_2 = e$, $\ell_1 = \ell_2 = \mu$ or $\ell_1 = e$ and $\ell_2 = \mu$.
For the first two cases, there is no special order for the four leptons. Recently, the branching
fractions of $\Jllll$ decays were calculated at the lowest order in nonrelativistic Quantum Chromodynamics~(NRQCD) factorization
in the SM~\cite{Chen:2020bju}. The smaller lepton mass would generate stronger collinear enhancement according to a recent explicit leading-order QED analysis~\cite{Chen:2020bju}.
Therefore, the predicted branching
fraction of $J/\psi\to e^+e^-e^+e^-$ is $(5.288\pm0.028)\times 10^{-5}$, significantly greater than that of $J/\psi\to e^+e^-\mu^+\mu^-$
($(3.763\pm0.020)\times 10^{-5}$) and two orders of magnitude greater than that of $J/\psi\to \mu^+\mu^-\mu^+\mu^-$~($(0.0974\pm0.0005)\times 10^{-5}$).
Therefore, the ratio $\mathcal{B}_{eeee}$: $\mathcal{B}_{ee\mu\mu}$: $\mathcal{B}_{\mu\mu\mu\mu}$ provides a good opportunity
to verify the validity of LFU, where $\mathcal{B}_{eeee}$, $\mathcal{B}_{ee\mu\mu}$ and $\mathcal{B}_{\mu\mu\mu\mu}$ represent the BFs of the $J/\psi\to e^+e^-e^+e^-$, $J/\psi\to e^+e^-\mu^+\mu^-$ and $J/\psi\to \mu^+\mu^-\mu^+\mu^-$ decays, respectively.

In addition, to date, the strength of charge-parity~(CP) violation in the quark-sector weak interaction has been found to be insufficient
to account for the observed matter–antimatter asymmetry in the universe, motivating further searches
for CP violation. Any new CP violation mechanism is usually constrained by the neutron electric dipole moment (nEDM)~\cite{Purcell:1950zz, Dar:2000tn,Afach:2015sja}. Recently, Sanchez-Puertas~\cite{Sanchez-Puertas:2018tnp} proposed a new test for CP
violation in the electromagnetic decay $\eta\to \mu^+\mu^- e^+e^-$, in which the sources of CP violation
are derived from dimension-six terms in the Standard Model Effective Field Theory~\cite{Grzadkowski:2010es}. This purely leptonic decay avoids the
strong constraints from the nEDM and could be studied at the proposed $\eta$ facility experiment REDTOP~\cite{Gatto:2016rae}.
Similarly, a test can be performed in leptonic decays of $\Jllll$ at BESIII, especially for the decay of $J/\psi\to e^+e^-\mu^+\mu^-$.

Using the world's largest $\psip$ data samples collected with the
BESIII detector~\cite{besiii}, we search for the decays $\Jllll$ with
$J/\psi$ events from $\Pppj$~\cite{Pppjpsi}.
Because of the enormous QED background from the two photon process,
$e^+ e^- \to \gamma^* \gamma^* \to \ell_1^+ \ell_1^- \ell_2^+ \ell_2^-$, this
analysis is performed using $J/\psi$ events from $\psip$ decays instead of
those from the larger $J/\psi$ data set.  Although the total number of
$J/\psi$ events from $\psip$ decays is one order of magnitude smaller, the QED
background can be suppressed to a negligible level by the requirement that
the $\pi^+ \pi^-$ recoil mass is near the $J/\psi$ mass.

This paper reports the first measurement of the branching fractions of
the decays $\Jllll$, which can be compared with theoretical
calculations.  The asymmetries of a CP observable constructed by a T-odd
triple-product~\cite{Gronau:2011cf,CPasymmetry,Shi:2019vus,Kang:2009iy,Kang:2010td} are presented for the first two channels~(
$J/\psi \to e^+ e^- e^+ e^-$ and $J/\psi \to e^+ e^- \mu^+ \mu^-$).

Details about the design and performance of the BESIII
detector are given in Refs.~\cite{besiii, BESIII2, BESIII3}.
Monte Carlo (MC) simulated data samples produced with a {\sc Geant}4-based~\cite{geant4} software package, which includes the geometric description of the BESIII detector and the detector
 response, are used to determine the detection efficiencies and to estimate backgrounds. The simulation includes the beam energy spread and initial state radiation in the $e^+ e^-$ annihilations modeled with the generator {\sc kkmc}~\cite{kkmc1, kkmc2}.
The data sample consists of $(448.1 \pm 2.9) \times 10^6$ $\psip$
events~\cite{psipnumber} collected with the BESIII
detector. Comparable amounts of inclusive MC simulated events
are used to study the backgrounds from $\psip$ decays.  The
production of the $\psip$ resonance is simulated by the MC event
generator {\sc kkmc}~\cite{kkmc1, kkmc2}.  The known decay modes are
generated by {\sc evtgen}~\cite{besevtgen1, besevtgen2} with
branching fractions taken from the Particle Data
Group~(PDG)~\cite{brpsip}, and the remaining unknown decays are
generated with the {\sc lundcharm} model~\cite{lundcharm1,
  lundcharm2}.  For the signal MC, the $\Pppj$ channel is generated
by {\sc jpipi}~\cite{besevtgen1, besevtgen2}, and the simulation of
the $\Jllll$ decay includes the polarization of the $J/\psi$ and the
angular distributions of the final states.  Since the $\pi^+\pi^-$
system is dominated by the $S-$wave~\cite{Pppjpsi, Ablikim:2006bz}, the
produced $J/\psi$ fully inherits the state of the $\psi(3686)$
polarization.

Charged particle tracks in the polar angle range $|\!\cos\theta|<0.93$
are reconstructed from hits in the main drift chamber~(MDC).  Tracks
with their point of closest approach to the beam line within $\pm10$
cm of the interaction point~(IP) in the beam direction, and within
1~cm in the plane perpendicular to the beam, are selected. At least
six charged tracks fulfilling these criteria are required.  The time-of-flight and specific energy loss ($dE/dx$) information are used
to calculate particle identification~(PID) probabilities~($prob$) for
the electron, pion, muon and kaon hypotheses. A track is considered to
be an electron if it satisfies $prob~(e)>prob~(\pi)$ and
$prob~(e)>prob~(K)$. For the two channels with muons in the final
states, muon candidates must satisfy $prob~(\mu)>prob~(e)$ and
$prob~(\mu)>prob~(K)$, and the deposited energy in the calorimeter~($E_{\rm
  EMC}$) for the muon candidate must be in the range of
[0.1,~0.3]~GeV. Any track not identified as an electron or muon is
assigned as a pion.

\begin{figure}[htbp]
\centering
  \includegraphics[width=1.02\linewidth]{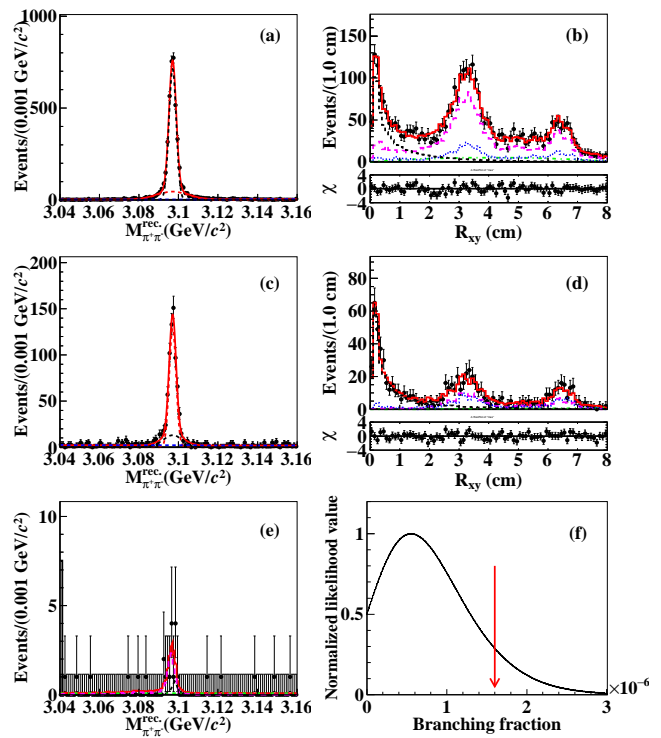}

  \caption{\label{Jpsito4e_fitdata} (a) Fit to the
  $M_{\pi^+\pi^-}^{\rm rec.}$ distribution of the $\psip$ data using a
  double Gaussian function for $\Jeeee$ channel. (b) Fit to the
  $R_{xy}$ distribution of for $\Jeeee$ channel. The black error
  bars represent data; the red solid curve indicates the overall fit;
  the black dashed line shows the signal; the blue dotted line is the
  inclusive MC background; the pink dashed line is the $J/\psi \to
  \gamma e^+ e^-$ background and the green dash-dotted line is the
  continuum background. (c) Fit to the $M_{\pi^+\pi^-}^{\rm rec.}$
  distribution using a double Gaussian function for the $\Jeemm$
  channel. (d) Fit to the $R_{xy}$ distribution for the
  $\Jeemm$ channel. (e) Fit to the $M_{\pi^+\pi^-}^{\rm rec.}$
  distribution for the $\Jmmmm$ channel. (f) Likelihood
  distribution versus the branching fraction of the $J/\psi \to \mu^+
  \mu^- \mu^+ \mu^-$ decay mode. The red arrow indicates the UL at the 90\% confidence level.}
\end{figure}

To identify $\pi^+\pi^- J/\psi$ candidates, two oppositely charged
tracks with momentum less than 0.45~GeV/$c$ are required and are used
to calculate the mass recoiling against them, $M^{\rm
  rec.}_{\pi^+\pi^-}$.  Using double Gaussian functions, fits are
  performed on the $M^{\rm rec.}_{\pi^+\pi^-}$ distributions of data,
  as shown in Fig.~\ref{Jpsito4e_fitdata} (a) and (c). The
  combinations with $M^{\rm rec.}_{\pi^+\pi^-}$ within a 5$\sigma$
  mass window around the $J/\psi$ peak are retained, where $\sigma$ is
  the resolution of the fitted double Gaussian function~($\sigma$ =
  $4.0\pm0.2$~MeV/$c^2$ for the $J/\psi \to e^+ e^- e^+ e^-$ channel; $\sigma$ =
  $2.8\pm0.5$~MeV/$c^2$ for the $J/\psi \to e^+ e^- \mu^+ \mu^-$ channel).  An
  energy-momentum constraint~(4C) kinematic fit is performed on the
  selected ~$\Pppj,$ $~\Jllll$ candidate events. For the channel
  $\Jeeee$, if more than one combination satisfies the 4C fit, the
  combination with the smallest $\chi^2_{\mathrm{4C}}$ is
  retained. For the last two channels, the combination
  with $M^{\rm
  rec.}_{\pi^{+}\pi^{-}}$ closest to the nominal $J/\psi$ mass is
selected. The $\chi^2_{\mathrm{4C}}$ values of the candidate events are
required to be less than 200.

To remove the $\Jeepp$ and $\Jpppp$ backgrounds, additional 4C
kinematic fits are performed, and events with $\chi^{2}_{\rm
  4C}(\Jeemm)<\chi^{2}_{\rm 4C}(\Jeepp)$ and $\chi^{2}_{\rm
  4C}(\Jmmmm)<\chi^{2}_{\rm 4C}(\Jpppp)$ are retained for the last two channels, respectively. To remove possible
contamination from the backgrounds with six charged tracks in the
final state, four of which are from $J/\psi$ decays, such as
$J/\psi\to 2(\pi^+\pi^-)$, $J/\psi\to 2(\pi K)$, and $J/\psi\to 2(K^+
K^-)$, \emph{etc.}, we apply further muon PID requirements for the
$\Jmmmm$ decay channel. The selection criteria of the depth in the muon counters of the muon
candidate are the same as Ref.~\cite{mupid}.

Possible background contributions are studied with data taken at
$\sqrt s = 3.773$~GeV~\cite{data_3650} and the $\psip$ inclusive MC
sample. The former indicates that the QED background is negligible.
Examination of the latter with an event topology analysis tool, TopoAna~\cite{xingyu:topology},
shows that the dominant backgrounds are from the channels $\Jgee$
and $J/\psi \to \gamma_\text{FSR}~e^+ e^-$ for the $\Jeeee$ decay,
$\Jgmm$ and $J/\psi \to \gamma_\text{FSR}~\mu^+ \mu^-$ for $\Jeemm$ and
$\Jmmmm$ decay modes, where $\gamma_\text{FSR}$ is a photon from final state
radiation which converts to an $e^+e^-$ pair in the detector material.
For the $\Jmmmm$ channel, two background events from
$\Jpppp$ in the inclusive MC sample survive the above selection criteria
and are considered as peaking background.

A photon-conversion finder~\cite{Rxyconv} is used to reconstruct the
photon conversion point. The distance from the IP to the reconstructed
conversion point of the lower momentum $e^+e^-$ pair, $R_{xy}$, is
used to separate the signal events from the photon conversion
background events.  For the $\Jeeee$ and $\Jeemm$ signal events, the
$R_{xy}$ distribution accumulates around 0~cm, because the
$e^+e^-$ pair comes directly from the $J/\psi$ decay. For the photon
conversion background, which is usually associated with the lower
momentum $e^+e^-$ pair, most of the photon conversions occur in the
beam pipe and inner wall of the MDC, and the $R_{xy}$ distribution
accumulates beyond 2~cm. However, there can be some contamination
under the peak at $R_{xy} = 0$ from events where another $e^+e^-$
pair occurs from photon conversion. Usually the conversion pairs in
these events have similar momentum with the lower momentum pair.  In
order to remove this background, the differences of the momenta
between the two electrons and the two positrons in the final state
are required to be greater than 1.0~GeV/$c$, i.e. $\Delta p
=(p_{e^{\pm}_h} - p_{e^{\pm}_l})>1.0$~GeV/$c$~($e^{\pm}_l$: $e^{\pm}$
with lower momentum, $e^{\pm}_h$: $e^{\pm}$ with higher momentum).

In order to determine the signal yields, unbinned maximum likelihood
fits are performed on the $R_{xy}$ distributions for the first two decay modes and to the $\pi^+\pi^-$ recoil mass~($M_{\pi^+\pi^-}^{\rm rec.}$) spectrum
for the $\Jmmmm$ channel, as shown in
Fig.~\ref{Jpsito4e_fitdata}~(b), (d) and (e), respectively. In the
fits to the $R_{xy}$ distributions for the first two channels, the signal is described by signal MC shape, while the
background is described by a combination of the inclusive MC, the
exclusive $J/\psi \to \gamma e^+ e^-$~($J/\psi \to \gamma \mu^+ \mu^-$) MC, and a 2$^{\rm nd}$ order
polynomial function. For the $\Jmmmm$ channel, the signal is
described by a MC-simulated shape convolved with a Gaussian
function, while the background is described by a combination of the
$\Jpppp$ MC shape and a 1$^{\rm st}$ order polynomial function. The
number of events of the peaking background from $\Jpppp$ is fixed to
the value estimated using the world average branching
fraction~\cite{brpsip}. The fitted signal yields, detection
efficiencies and measured branching fractions are listed in
Table~\ref{Jpsito4e fit results}. The measured branching fractions are
determined by
\begin{equation}
\label{Jpsito4e formulaBR}
{\mathcal B}(\Jllll)\!=\!\frac{N_{\rm sig}}{N_{\psip}\!\cdot\!\epsilon\!\cdot\! \mathcal{B}(\psip)},
\end{equation}
where $N_{\rm sig}$ is the number of observed signal events,
$N_{\psip}$ is the number of $\psip$ events~\cite{psipnumber},
and $\epsilon$ is the detection efficiency determined by MC
simulation. The branching fraction $\mathcal{B}(\psip) \equiv \mathcal{B}(\psip \to \pi^+ \pi^-
J/\psi)$ is taken from the PDG~\cite{brpsip}.
To obtain a more accurate of detecting the decay of $\Jeeee$, the events of the signal MC sample are weighted according to the $e^{+}_{l}$ momentum versus $e^{-}_{l}$ momentum distribution in data. The weight factors are the $\epsilon_{\rm data}^{i}/\epsilon_{\rm MC}^{i}$ ratios which are obtained in different momentum bins, where $\epsilon_{\rm data}^{i}$ and $\epsilon_{\rm MC}^{i}$ are the efficiencies of $e^+e^-e^+e^-$ candidates in the $i$-th bin from data and MC simulation, respectively. The resultant detection efficiency is $(8.22\pm0.02)$\%.

\begin{table*}[htp]\centering\footnotesize
%\resizebox{15cm}{28pt}{
\caption{\label{Jpsito4e fit results} Observed signal yields, detection efficiencies, numbers of observed events with $C_{\rm T} > 0$ and $C_{\rm T} < 0$, measured branching fraction and $\mathcal{A}_{T}$ values for different decay modes. The first uncertainty is statistical and the second is systematic.}
\begin{spacing}{1.2}
\begin{tabular}{c r@{$\,\pm\,$}l r@{$\,\pm\,$}l r@{$\,\pm\,$}l r@{$\,\pm\,$}l ccc}
\hline\hline
  &\multicolumn{10}{c}{This work}  &Theory~\cite{Chen:2020bju}   \\
Decay mode   &\multicolumn{2}{c}{$N_{\rm sig}$}     &\multicolumn{2}{c}{$\epsilon(\%)$}  &\multicolumn{2}{c}{$N(C_{\rm T} > 0)$}  &\multicolumn{2}{c}{$N(C_{\rm T} < 0)$}     &${\mathcal B}(\times 10^{-5})$  &$\mathcal{A}_{T}$ &${\mathcal B}(\times 10^{-5})$ \\\hline
  $\Jeeee$     &700  &39    &8.22  &0.02   &355  &27    &363  &28         &$5.48\pm0.31\pm0.45$  &$-0.012\pm0.054\pm0.010$  &$5.288\pm0.028$ \\
  $\Jeemm$     &354  &22    &6.46  &0.04   &193   &15    &170   &15         &$3.53\pm0.22\pm0.13$  &$0.062\pm0.059\pm0.006$  &$3.763\pm0.020$ \\
  $\Jmmmm$     &3.4  &4.1     &3.96  &0.03    &\multicolumn{2}{c}{- ~~}  &\multicolumn{2}{c}{-~~ }       &$<0.16$ &-  &$0.0974\pm0.0005$ \\
\hline\hline
\end{tabular}
\end{spacing}
%}
\end{table*}

The following sources of systematic uncertainty for the
branching fraction measurement are considered: tracking
efficiency, PID efficiency, total number of $\psip$ events, branching
fraction of $\Pppj$, kinematic fit, $\pi^{+}\pi^{-}$ recoil mass
window, fit range, signal PDF shape, background PDF shape, and the
$\Delta p$ cut.

To study the systematic uncertainties related to the tracking and PID
efficiencies of electrons and positrons, control samples of radiative
Bhabha events are selected, \emph{i.e.} $\eegee$, at the $\psip$ energy for both the experimental data and MC simulation. The differences between
the data and MC simulation in different momentum and polar angle
regions are obtained from the control samples. The systematic
uncertainties are calculated according to the
momentum and polar angle distributions in the signal MC sample.  They are
assigned as 1.1\% and 0.3\% from tracking efficiencies, 2.0\% and 1.7\% from PID efficiencies of the first two channels, respectively.

The systematic uncertainties from the tracking and PID efficiencies of
muons are estimated through a control sample of $e^+ e^- \to
(\gamma)~\mu^+\mu^-$~\cite{mupid}. The uncertainties from the tracking
and PID efficiencies of each muon track are 0.1\% and 0.8\%,
respectively. The systematic uncertainties from the pion tracking
efficiency are estimated through a control sample of $\Pppj$, which
gives 1\% for each pion track~\cite{trackingpi}. The systematic
uncertainty on the number of $\psip$ events is
0.6\%~\cite{psipnumber}, and the uncertainty of the branching fraction of
$\Pppj$ is 0.9\%~\cite{brpsip}.

In the 4C kinematic fits, the helix parameters of simulated charged tracks are
corrected to reduce the discrepancy between data and MC simulation as
described in Ref.~\cite{helix}. The correction factors are obtained
by studying a control sample of $\psip\to\pi^+\pi^-J/\psi, J/\psi\to
\ell^+\ell^-$. The MC samples with the corrected track helix parameters are
taken as the nominal ones. The differences between detection
efficiencies obtained from MC samples with and without the correction
are taken as the uncertainties,
which are 1.0\%, 0.7\% and 0.9\% for the three channels, respectively.

The systematic uncertainties due to the maximum likelihood fits are
studied as follows: For the first two channels, the signal shape
 is replaced with the signal MC convolved with a gaussian function, and the uncertainties from the signal shape are evaluated to be 1.7\% and 1.2\%, respectively.
The 2$^{\rm nd}$ order polynomial function is changed to a 3$^{\rm rd}$
order one to estimate the uncertainties from the background
PDF shape, which are assigned to be 0.8\% and 0.1\%, respectively. To evaluate the systematic effects from the requirement on
the recoil mass of $\pi^+\pi^-$, an alternative shape from the
signal MC sample is used. The relative differences in the number of signal events, 0.4\% and 0.1\%, respectively, are taken as the systematic uncertainties. To estimate the systematic
uncertainties from the fit range, the range of $R_{xy}$ is
changed from $[0.0, 8.0]$ to $[0.0, 9.0]$, and the changes of the measured
branching fractions are taken as the systematic uncertainties, which are evaluated to be 1.0\% and 0.1\%, respectively. The
systematic uncertainty from the selection on $\Delta p$ is
estimated by using a different requirement~($(p_{e^{\pm}_h} -
p_{e^{\pm}_l})>1.10$~GeV/$c$) on this variable, set to be 3.0\%.
To obtain reliable signal efficiency, the signal MC
sample $\Jeeee$ is weighted to match the data. To estimate the associated systematic uncertainty, the weight factors are randomly changed within one standard deviation in each bin one thousand times to re-obtain the signal efficiency. The distribution of the resulting signal efficiency is fit with a Gaussian function, and the standard deviation of 6.5\% is taken as the systematic uncertainty.
 For the $\Jmmmm$
mode, the $\pi^+\pi^-$ recoil mass window is changed from $[3.04,
  3.16]$~GeV/$c^2$ to $[3.06, 3.16]$~GeV/$c^2$, and 0.9\% is taken as the systematic uncertainty from this source. The 1$^{\rm
  st}$ order polynomial function is changed to a 2$^{\rm nd}$ order
one to estimate the uncertainty from the background PDF shape, which is
assigned to be 0.3\%. In
estimating the uncertainty from the background events of $J/\psi \to
\pi^+ \pi^- \pi^+ \pi^-$, the fits are performed twenty thousand times
with the number of the background events
sampled from a Gaussian distribution based on its statistical error. The distribution of
the relative difference on the signal yield is fitted by a Gaussian
function, and its width of 0.2\% is assigned as the systematic uncertainty
from this source. The total systematic uncertainties of the BF measurement are 8.2\%, 3.6\%, and 4.2\% for the three channels, respectively, obtained by adding the above effects quadratically.

As indicated in Fig.~\ref{Jpsito4e_fitdata}~(e), no significant signal
is observed for the $\Jmmmm$ channel, so the branching fraction
upper limit~(UL) is determined according to Eq.~\ref{Jpsito4e formulaBR} based on a Bayesian method~\cite{Bayesian}.
 A series of fits of the $M_{\pi^+\pi^-}^{\rm rec.}$ distribution are
carried out fixing the BF at different values, and the
resultant curve of likelihood values as a function of the
BF is convolved with a Gaussian function to take into account the
overall systematic uncertainty, as shown in Fig.~\ref{Jpsito4e_fitdata}~(f).
The UL on the BF is obtained when the integral of the
likelihood curve in the positive domain reaches 90\% of its total value.

Following Refs.~\cite{Gronau:2011cf,CPasymmetry}, we measure
correlations between the final state leptons. In the rest
frame of $J/\psi$, we define $C_{\rm T} = (\vec{p}_{\ell_{a}^{+}}\times
\vec{p}_{\ell_{a}^{-}}) \cdot \vec{p}_{\ell_{b}^{+}}$, where $\ell_a$ and $\ell_b$
are final-state leptons: $\ell_a = \ell_b = e$ for $\Jeeee$, while $\ell_a = e$
and $\ell_b = \mu$ for $\Jeemm$.  In the former case, $\ell_a^+$ and $\ell_a^-$
are taken as the leptons with lower momenta, and $\ell_b^+$ those with higher
momenta~\cite{Gronau:2011cf,Aaij:2016cla}. A CP asymmetry observable
based on CPT invariance~\cite{Gronau:2011cf,CPasymmetry},
$\mathcal{A}_{T}$, can be constructed with the number of events of
positive and negative $C_{\rm T}$ values:
\begin{equation}
\label{Jpsito4e AT}
\mathcal{A}_{T} = \frac{N(C_{\rm T} > 0) - N(C_{\rm T} < 0)}{N(C_{\rm T} > 0) + N(C_{\rm T} < 0)},
\end{equation}

The numbers of events with $C_{\rm T} > 0$ and $C_{\rm
  T} < 0$ are evaluated by a simultaneous fit to the $R_{xy}$
distributions of data using the same fit method as that used in
the branching fraction measurements. The parameters of the signal shape are shared in the fits, while the other parameters are free.
Taking into account the contributions of the same systematic uncertainties
as in the branching fraction measurement, the respective numbers of events and measured
asymmetries of the $\Jeeee$ and $\Jeemm$ channels are listed in Table~\ref{Jpsito4e fit results}.

In summary, based on a data sample of $4.481\times 10^8$ $\psip$
events collected with the BESIII detector, the decay processes of
$\Jllll$ are investigated.  The decay channels of $\Jeeee$ and
$\Jeemm$ are observed for the first time, and the corresponding
branching fractions are measured. The theoretical predictions based on NRQCD~\cite{Chen:2020bju} are consistent with our measurements of the branching fractions for these two channels.
No signal is observed for the
$\Jmmmm$ channel, and the UL of the branching fraction is
determined.
No CP violation is observed. All measured results are summarized in Table~\ref{Jpsito4e fit results}.
The ratio $\mathcal{B}_{eeee}$/$\mathcal{B}_{ee\mu\mu}$
is calculated to be $(1.55\pm0.13\pm0.14)$, which agrees with theory
within 1$\sigma$~\cite{Chen:2020bju}. The UL on the ratio
$\mathcal{B}_{\mu\mu\mu\mu}$/$\mathcal{B}_{eeee}$ is calculated with
\begin{equation}
\label{ratio}
\frac{\mathcal{B}_{\mu\mu\mu\mu}}{\mathcal{B}_{eeee}}<\frac{N_{\mathrm{UL}}^{4\mu} / \epsilon^{4\mu}}{N^{4e} / \epsilon^{4e}} \frac{1}{\left(1-\sigma_{13}\right)},
\end{equation}
Here $N_{\mathrm{UL}}^{4\mu}$ is the 90\% UL on the number of
observed events for $\Jmmmm$ decay; $\epsilon^{4\mu}$ is the
MC-determined efficiency for the channel; $N^{4e}$ is the number of
events for the $\Jeeee$ decay; $\epsilon^{4e}$ is the MC-determined
efficiency for this channel; $\sigma_{13} =
\sqrt{(\sigma_{13}^{\rm stat})^2 + (\sigma_{13}^{\rm sys})^2} = 10.9\%$, where
$\sigma_{13}^{\rm stat}$ is the relative statistical error of $N^{4e}$
(5.7\%) and $\sigma_{13}^{\rm sys}$ is the total relative systematic error
for $\Jeeee$ and $\Jmmmm$ decay channels. The UL on the ratio
$\mathcal{B} _{\mu\mu\mu\mu}$/$\mathcal{B}_{eeee}$ is determined to be
0.033. Similarly, the UL on the ratio
$\mathcal{B}_{\mu\mu\mu\mu}$/$\mathcal{B}_{ee\mu\mu}$ is determined to
be 0.050.

The BESIII collaboration thanks the staff of BEPCII and the IHEP computing center for their strong support.
The authors would like to extend thanks to Prof. Xian-Wei Kang for useful discussions and helpful advice.
This work is supported in part by National Key Research and Development Program of China under Contracts Nos. 2020YFA0406300, 2020YFA0406400; National Natural Science Foundation of China (NSFC) under Contracts Nos. 11805037, 11975118, 11575077, 12165022; the Natural Science Foundation of Hunan Province under Contract Nos. 2020RC3054 and 2019JJ30019;
the Chinese Academy of Sciences (CAS) Large-Scale Scientific Facility Program; Joint Large-Scale Scientific Facility Funds of the NSFC and CAS under Contracts Nos. U1832121,
U1732263, U1832207; CAS Key Research Program of Frontier Sciences under Contract No. QYZDJ-SSW-SLH040; 100 Talents Program of CAS; INPAC and Shanghai Key Laboratory for Particle Physics and Cosmology; Yunnan Fundamental Research Project under Contract No. 202301AT070162;
ERC under Contract No. 758462; European Union Horizon 2020 research and innovation programme under Contract No. Marie Sklodowska-Curie grant agreement No 894790; German Research Foundation DFG under Contracts Nos. 443159800, Collaborative Research Center CRC 1044, FOR 2359, GRK 214; Istituto Nazionale di Fisica Nucleare, Italy; Ministry of Development of Turkey under Contract No. DPT2006K-120470; National Science and Technology fund; Olle Engkvist Foundation under Contract No. 200-0605; STFC (United Kingdom); The Knut and Alice Wallenberg Foundation (Sweden) under Contract No. 2016.0157; The Royal Society, UK under Contracts Nos. DH140054, DH160214; The Swedish Research Council; U. S. Department of Energy under Contracts Nos. DE-FG02-05ER41374, DE-SC-0012069.

\end{document}

%% file: BESIII_author.tex
\author{
M.~Ablikim$^{1}$, M.~N.~Achasov$^{10,c}$, P.~Adlarson$^{67}$, S. ~Ahmed$^{15}$, M.~Albrecht$^{4}$, R.~Aliberti$^{28}$, A.~Amoroso$^{66A,66C}$, M.~R.~An$^{32}$, Q.~An$^{63,49}$, X.~H.~Bai$^{57}$
% ...
\\
\vspace{0.2cm}
(BESIII Collaboration)\\
\vspace{0.2cm} {\it
$^{1}$ Institute of High Energy Physics, Beijing 100049, People's Republic of China\\
% ...
$^{4}$ Bochum Ruhr-University, D-44780 Bochum, Germany\\
% ...
$^{10}$ G.I. Budker Institute of Nuclear Physics SB RAS (BINP), Novosibirsk 630090, Russia\\
% ...
$^{15}$ Helmholtz Institute Mainz, Johann-Joachim-Becher-Weg 45, D-55099 Mainz, Germany\\
% ...
$^{28}$ Johannes Gutenberg University of Mainz, Johann-Joachim-Becher-Weg 45, D-55099 Mainz, Germany\\
% ...
$^{32}$ Liaoning Normal University, Dalian 116029, People's Republic of China\\
% ...
$^{49}$ State Key Laboratory of Particle Detection and Electronics, Beijing 100049, Hefei 230026, People's Republic of China\\
% ...
$^{57}$ University of Jinan, Jinan 250022, People's Republic of China\\
% ...
$^{63}$ University of Science and Technology of China, Hefei 230026, People's Republic of China\\
% ...
$^{66}$ University of Turin and INFN, (A)University of Turin, I-10125, Turin, Italy; (B)University of Eastern Piedmont, I-15121, Alessandria, Italy; (C)INFN, I-10125, Turin, Italy\\
$^{67}$ Uppsala University, Box 516, SE-75120 Uppsala, Sweden\\
% ...
$^{c}$ Also at the Novosibirsk State University, Novosibirsk, 630090, Russia\\
% ...
}
}

%% file: jpsito4l.bbl
\begin{thebibliography}{**}

%\cite{Glashow:1959wxa}
\bibitem{Glashow:1959wxa}
S.~L.~Glashow,
%\emph{The renormalizability of vector meson interactions},
\href{https://doi.org/10.1016/0029-5582(59)90196-8}{Nucl. Phys. {\bf 10}, 107-117 (1959).}

\bibitem{Salam:1959zz}
A.~Salam and J.~C.~Ward,
%\emph{Weak and electromagnetic interactions},
\href{https://doi.org/10.1007/BF02726525}{Nuovo Cim. {\bf 11}, 568-577 (1959).}

%\cite{Weinberg:1967tq}
\bibitem{Weinberg:1967tq}
S.~Weinberg,
%\emph{A Model of Leptons},
\href{https://doi.org/10.1103/PhysRevLett.19.1264}{Phys. Rev. Lett. {\bf 19}, 1264 (1967).}
%13180 citations counted in INSPIRE as of 30 Apr 2021


%\cite{Aaij:2014ora}
\bibitem{Aaij:2014ora}
R.~Aaij \textit{et al.} [LHCb],
%\emph{Test of lepton universality using $B^{+}\rightarrow K^{+}\ell^{+}\ell^{-}$ decays},
\href{https://doi.org/10.1103/PhysRevLett.113.151601}{Phys. Rev. Lett. {\bf 113}, 151601 (2014).}
%1096 citations counted in INSPIRE as of 30 Apr 2021

%\cite{Bordone:2016gaq}
\bibitem{Bordone:2016gaq}
M.~Bordone, G.~Isidori and A.~Pattori,
%\emph{On the Standard Model predictions for $R_K$ and $R_{K^*}$},
\href{https://doi.org/10.1140/epjc/s10052-016-4274-7}{Eur. Phys. J. C \textbf{76}, 440 (2016).}
%324 citations counted in INSPIRE as of 30 Apr 2021

%\cite{Aaij:2019wad}
\bibitem{Aaij:2019wad}
R.~Aaij \textit{et al.} [LHCb],
%\emph{Search for lepton-universality violation in $B^+\to K^+\ell^+\ell^-$ decays},
\href{https://doi.org/10.1103/PhysRevLett.122.191801}{Phys. Rev. Lett. \textbf{122}, 191801 (2019).}
%332 citations counted in INSPIRE as of 30 Apr 2021

%\cite{Abdesselam:2019lab}
\bibitem{Abdesselam:2019lab}
S.~Choudhury \textit{et al.} [BELLE],
%\emph{Test of lepton flavor universality and search for lepton flavor violation in $B \rightarrow K\ell \ell$ decays},
\href{https://doi.org/10.1007/JHEP03(2021)105}{J. High Energ. Phys. \textbf{03}, 105 (2021).}
%68 citations counted in INSPIRE as of 30 Apr 2021

%\cite{Aaij:2021vac}
\bibitem{Aaij:2021vac}
R.~Aaij \textit{et al.} [LHCb],
%\emph{Test of lepton universality in beauty-quark decays},
\href{https://doi.org/10.1038/s41567-021-01478-8}{Nat. Phys. \textbf{18}, 277-282 (2022).}
%41 citations counted in INSPIRE as of 30 Apr 2021


%\cite{Abi:2021gix}
\bibitem{Abi:2021gix}
B.~Abi \textit{et al.} [Muon g-2],
%\emph{Measurement of the Positive Muon Anomalous Magnetic Moment to 0.46~ppm},
\href{https://doi.org/10.1103/PhysRevLett.126.141801}{Phys. Rev. Lett. \textbf{126}, 141801 (2021).}
%87 citations counted in INSPIRE as of 01 May 2021

%\cite{Bennett:2006fi}
\bibitem{Bennett:2006fi}
G.~W.~Bennett \textit{et al.} [Muon g-2],
%\emph{Final Report of the Muon E821 Anomalous Magnetic Moment Measurement at BNL},
\href{https://doi.org/10.1103/PhysRevD.73.072003}{Phys. Rev. D \textbf{73}, 072003 (2006).}
%2424 citations counted in INSPIRE as of 01 May 2021


%\cite{Ban:2021tos}
\bibitem{Ban:2021tos}
K.~Ban, Y.~Jho, Y.~Kwon, S.~C.~Park, S.~Park and P.~Y.~Tseng,
%\emph{A comprehensive study of vector leptoquark on the $B$-meson and Muon g-2 anomalies},
\href{https://doi.org/10.1093/ptep/ptac159}{Prog. Theor. Exp. Phys. \textbf{2023}, 013B01 (2023).}
%1 citations counted in INSPIRE as of 03 May 2021

%\cite{Marzocca:2021azj}
\bibitem{Marzocca:2021azj}
D.~Marzocca and S.~Trifinopoulos,
%\emph{A Minimal Explanation of Flavour Anomalies: B-Meson Decays, Muon Magnetic Moment, and the Cabbibo Angle},
\href{https://doi.org/10.1103/PhysRevLett.127.061803}{Phys. Rev. Lett. \textbf{127}, 061803 (2021).}
%1 citations counted in INSPIRE as of 03 May 2021

%\cite{Borah:2021jzu}
\bibitem{Borah:2021jzu}
D.~Borah, M.~Dutta, S.~Mahapatra and N.~Sahu,
%\emph{Muon $(g-2)$ and XENON1T Excess with Boosted Dark Matter in $L_{\mu}-L_{\tau}$ Model},
\href{https://doi.org/10.1016/j.physletb.2021.136577}{Phys. Lett. B \textbf{820}, 136577 (2021).}
%2 citations counted in INSPIRE as of 03 May 2021

%\cite{Lancierini:2021sdf}
\bibitem{Lancierini:2021sdf}
D.~Lancierini, G.~Isidori, P.~Owen and N.~Serra,
%\emph{On the significance of new physics in $b\to s\ell^+\ell^-$ decays},'
\href{https://doi.org/10.1016/j.physletb.2021.136644}{Phys. Lett. B. \textbf{822}, 136644 (2021).}
%2 citations counted in INSPIRE as of 03 May 2021

%\cite{Du:2021zkq}
\bibitem{Du:2021zkq}
M.~Du, J.~Liang, Z.~Liu and V.~Tran,
%\emph{A vector leptoquark interpretation of the muon $g-2$ and $B$ anomalies},
\href{https://arxiv.org/abs/2104.05685v1}{arXiv:2104.05685.}
%2 citations counted in INSPIRE as of 03 May 2021

%\cite{Kawamura:2021ygg}
\bibitem{Kawamura:2021ygg}
J.~Kawamura and S.~Raby,
%\emph{$\ge 4 \mu$ signal from a vector-like lepton decaying to a muon-philic $Z^\prime$ boson at the LHC},
\href{https://doi.org/10.1103/PhysRevD.104.035007}{Phys. Rev. D \textbf{104}, 035007 (2021).}
%3 citations counted in INSPIRE as of 03 May 2021

%\cite{Chen:2021vzk}
\bibitem{Chen:2021vzk}
J.~Chen, Q.~Wen, F.~Xu and M.~Zhang,
%\emph{Flavor Anomalies Accommodated in A Flavor Gauged Two Higgs Doublet Model},
\href{https://arxiv.org/abs/2104.03699}{arXiv:2104.03699.}
%7 citations counted in INSPIRE as of 03 May 2021

%\cite{Yin:2021yqy}
\bibitem{Yin:2021yqy}
W.~Yin,
%\emph{Radiative lepton mass and muon $g-2$ with suppressed lepton flavor and CP violations},
\href{https://doi.org/10.1007/JHEP08(2021)043}{J. High Energ. Phys. \textbf{08}, 043 (2021).}
%7 citations counted in INSPIRE as of 03 May 2021

%\cite{Arcadi:2021cwg}
\bibitem{Arcadi:2021cwg}
G.~Arcadi, L.~Calibbi, M.~Fedele and F.~Mescia,
%\emph{Muon $g-2$ and $B$-anomalies from Dark Matter},
\href{https://doi.org/10.1103/PhysRevLett.127.061802}{Phys. Rev. Lett. \textbf{127}, 061802 (2021).}
%10 citations counted in INSPIRE as of 02 May 2021

%\cite{Baum:2021qzx}
\bibitem{Baum:2021qzx}
S.~Baum, M.~Carena, N.~R.~Shah and C.~E.~M.~Wagner,
%\emph{The Tiny (g-2) Muon Wobble from Small-$\mu$ Supersymmetry},
\href{https://doi.org/10.1007/JHEP01(2022)025}{J. High Energ. Phys. \textbf{01}, 025 (2022).}
%13 citations counted in INSPIRE as of 02 May 2021

%\cite{Cadeddu:2021dqx}
\bibitem{Cadeddu:2021dqx}
M.~Cadeddu, N.~Cargioli, F.~Dordei, C.~Giunti and E.~Picciau,
%\emph{Muon and electron g-2, proton and cesium weak charges implications on dark $\mathbf{Z_d}$ models},
\href{https://doi.org/10.1103/PhysRevD.104.L011701}{Phys. Rev. D \textbf{104}, 011701 (2021).}
%10 citations counted in INSPIRE as of 02 May 2021

%\cite{Ellis:2021ixr}
\bibitem{Ellis:2021ixr}
J.~Ellis,
%\emph{Following in Tini's Giant Footsteps},
\href{https://doi.org/10.5506/APhysPolB.52.561}{Acta Phys. Pol. B \textbf{52}, 561 (2021).}
%0 citations counted in INSPIRE as of 02 May 2021



\bibitem{bes3jpsi2ll} M. Ablikim \textit{et al.} (BES Collaboration),
\href{https://doi.org/10.1016/j.physletb.2016.08.011}{Phys. Lett. B {\bf 761}, 98 (2016).}

\bibitem{Chen:2020bju}
W.~Chen, Y.~Jia, Z.~Mo, J.~Pan and X.~Xiong,
\href{https://doi.org/10.1103/PhysRevD.104.094023}{Phys. Rev. D {\bf 104}, 094023 (2021).}

\bibitem{Purcell:1950zz}
E.~M.~Purcell and N.~F.~Ramsey,
\href{https://doi.org/10.1103/PhysRev.78.807}{Phys. Rev. {\bf 78}, 807 (1950).}

\bibitem{Dar:2000tn}
S.~Dar,
\href{https://arxiv.org/abs/hep-ph/0008248}{arXiv:hep-ph/0008248.}

\bibitem{Afach:2015sja}
J.~M.~Pendlebury, S.~Afach, N.~J.~Ayres, C.~A.~Baker, G.~Ban, G.~Bison, K.~Bodek, M.~Burghoff, P.~Geltenbort and K.~Green \textit{et al.},
\href{https://doi.org/10.1103/PhysRevD.92.092003}{Phys. Rev. D {\bf 92}, 092003 (2015).}

\bibitem{Sanchez-Puertas:2018tnp}
P.~Sanchez-Puertas,
\href{https://doi.org/10.1007/JHEP01(2019)031}{J. High Energ. Phys. {\bf 01}, 031 (2019).}

\bibitem{Grzadkowski:2010es}
B.~Grzadkowski, M.~Iskrzynski, M.~Misiak and J.~Rosiek,
\href{https://doi.org/10.1007/JHEP10(2010)085}{J. High Energ. Phys. {\bf 10}, 085 (2010).}

\bibitem{Gatto:2016rae}
C.~Gatto \textit{et al}. [REDTOP],
\href{https://doi.org/10.22323/1.282.0812}{Proceedings of Science {\bf ICHEP2016}, 812 (2016).}

\bibitem{besiii} M. Ablikim \textit{et al}. (BESIII Collaboration),
\href{https://doi.org/10.1016/j.nima.2009.12.050}{Nucl. Instrum. Methods Phys. Res., Sect. A {\bf 614}, 345 (2010).}

\bibitem{Pppjpsi} J. Z. Bai \textit{et al.} (BES Collaboration),
\href{https://doi.org/10.1103/PhysRevD.62.032002}{Phys. Rev. D {\bf 62}, 032002 (2000).}


  \bibitem{Gronau:2011cf} M. Gronau and J. L. Rosner,
  \href{https://doi.org/10.1103/PhysRevD.84.096013}{Phys. Rev. D {\bf 84}, 096013 (2011).}

\bibitem{CPasymmetry} I. I. Bigi, X. W. Kang, H. B. Li,
  \href{https://doi.org/10.1088/1674-1137/42/1/013101}{Chin. Phys. C {\bf 42}, 013101~(2018).}

\bibitem{Shi:2019vus}
X.~D.~Shi, X.~W.~Kang, I.~Bigi, W.~P.~Wang and H.~P.~Peng,
%``Prospects for CP and P violation in $\Lambda_{c}^+$ decays at Super Tau Charm Facility,''
\href{https://doi.org/10.1103/PhysRevD.100.113002}{Phys. Rev. D \textbf{100}, 113002 (2019).}

\bibitem{Kang:2009iy}
X.~W.~Kang and H.~B.~Li,
%``Study of CP violation in D ---\ensuremath{>} VV decay at BESIII,''
\href{https://doi.org/10.1016/j.physletb.2010.01.024}{Phys. Lett. B \textbf{684}, 137 (2010).}

\bibitem{Kang:2010td}
X.~W.~Kang, H.~B.~Li, G.~R.~Lu and A.~Datta,
%``Study of CP violation in $\Lambda_c^+$ decay,''
\href{https://doi.org/10.1142/S0217751X11053432}{Int. J. Mod. Phys. A \textbf{26}, 2523-2535 (2011).}
%[arXiv:1003.5494 [hep-ph]].

\bibitem{BESIII2} C. H. Yu \textit{et al.},
\href{http://ir.ihep.ac.cn/handle/311005/247347}{Proceedings of IPAC 2016,~(2016).}

\bibitem{BESIII3} M. Ablikim \textit{et al.} (BESIII Collaboration),
\href{https://doi.org/10.1088/1674-1137/44/4/040001}{Chin. Phys. C {\bf 44}, 040001~(2020).}

\bibitem{geant4} S. Agostinelli et al. (GEANT4 Collaboration),
\href{https://doi.org/10.1016/S0168-9002(03)01368-8}{Nucl. Instrum. Meth. A {\bf 506}, 250~(2003).}

  \bibitem{kkmc1} S. Jadach, B. F. L. Ward and Z. Was,
  \href{https://doi.org/10.1016/S0010-4655(00)00048-5}{Comput. Phys. Commun. {\bf 130}, 260 (2000).}

  \bibitem{kkmc2} S. Jadach, B. F. L. Ward and Z. Was,
  \href{https://doi.org/10.1103/PhysRevD.63.113009}{Phys. Rev. D {\bf 63}, 113009 (2001).}

\bibitem{psipnumber} M. Ablikim \textit{et al.} (BESIII Collaboration),
\href{https://doi.org/10.1088/1674-1137/42/2/023001}{Chin. Phys.C {\bf 42}, 023001~(2018).}

  \bibitem{besevtgen1} R.~G. Ping,
  \href{https://doi.org/10.1088/1674-1137/32/8/001}{Chin. Phys. C {\bf 32}, 599 (2008).}

  \bibitem{besevtgen2} D. J. Lange,
  \href{https://doi.org/10.1016/S0168-9002(01)00089-4}{Nucl. Instrum. Meth., Sect. A {\bf 462}, 152 (2001).}

  \bibitem{brpsip} P. A. Zyla \textit{et al.} (Particle Data Group),
  \href{https://academic.oup.com/ptep/article/2020/8/083C01/5891211}{Prog. Theor. Exp. Phys. {\bf 2020}, 083C01 (2020).}


  \bibitem{lundcharm1} J. C. Chen, G. S. Huang, X. R. Qi, D. H. Zhang, and Y. S. Zhu,
  \href{https://doi.org/10.1103/PhysRevD.62.034003}{Phys. Rev. D {\bf 62}, 034003~(2000).}

  \bibitem{lundcharm2} R. L. Yang, R. G. Pong, and H. Chen,
  \href{https://doi.org/10.1088/0256-307X/31/6/061301}{Chin. Phys. Lett. {\bf 31}, 061301~(2014).}


  \bibitem{Ablikim:2006bz} M.~Ablikim \textit{et al.} (BES Collaboration),
  \href{https://doi.org/10.1016/j.physletb.2006.11.056}{Phys. Lett. B \textbf{645}, 1~(2007).}

  \bibitem{mupid} M. Ablikim \textit{et al.} (BESIII Collaboration),
  \href{https://doi.org/10.1103/PhysRevD.89.051104}{Phys. Rev. D {\bf 89}, 051104~(2014).}

  \bibitem{data_3650} M. Ablikim \textit{et al.} (BESIII Collaboration),
  \href{https://doi.org/10.1088/1674-1137/37/12/123001}{Chin. Phys. C {\bf 37}, 123001~(2013).}

  \bibitem{xingyu:topology} X. Zhou, S. Du, G. Li, C. Shen,
  %\emph{TopoAna: A generic tool for the event type analysis of inclusive Monte-Carlo samples in high energy physics experiments},
  \href{https://doi.org/10.1016/j.cpc.2020.107540}{Comput. Phys. Commun. {\bf 258}, 107540 (2021).}

   \bibitem{Rxyconv} Z. R. Xu and K. L. He,
   \href{https://doi.org/10.1088/1674-1137/36/8/010}{Chin. Phys. C {\bf 36}, 742~(2012).}

  %\bibitem{CB_func} M. J. Oreglia, Ph.D Thesis, SLAC-236, Appendix D (1980); J. E. Gaiser, Ph.D Thesis,
%SLAC-255, Appendix F, (1982); T. Skwarnicki, Ph.D Thesis, DESY F31-86-02, Appendix E (1986).

  \bibitem{trackingpi} M. Ablikim \textit{et al.} (BESIII Collobarotion),
  \href{https://journals.aps.org/prd/pdf/10.1103/PhysRevD.83.112005}{Phys. Rev. D {\bf 83}, 112005 (2011).}

  \bibitem{helix} M. Ablikim \textit{et al.} (BESIII Collaboration),
   \href{https://journals.aps.org/prd/pdf/10.1103/PhysRevD.87.012002}{Phys. Rev. D {\bf 87}, 012002 (2013).}

  %\bibitem{landau} L. Landau, J. Phys. USSR {\bf 8}, 201~(1944);
  %W. W. M. Allison and J. H. Cobb,
  %\href{https://www.annualreviews.org/doi/pdf/10.1146/annurev.ns.30.120180.001345}{Annu. Rev. Nucl. Part. Sci. {\bf 30}, 253~(1980).}

  \bibitem{Bayesian} Y. S. Zhu,\\
  %\emph{Upper limit for Poisson variable incorporating systematic uncertainties by Bayesian approach},
  \href{https://doi.org/10.1016/j.nima.2007.05.116}{Nucl. Instrum. Methods Phys. Res.,
  Sect. A {\bf 578}, 322 (2007).}


  \bibitem{Aaij:2016cla} R. Aaij \textit{et al.} (LHCb Collaboration),
  \href{https://doi.org/10.1038/nphys4021}{Nature Phys. {\bf 13}, 391-396~(2017).}

\end{thebibliography}
